\documentclass[aps,prb,twocolumn,showpacs,floatfix,preprintnumbers,superscriptaddress,amsmath,amssymb,footinbib]{revtex4-1}

\usepackage{graphicx}
\usepackage{dcolumn}
\usepackage{bm}

\newcommand{\sbr}[1]{\left[#1\right]}

\begin{document}

\preprint{bsConf v3.5 arXiv}

\title{Bulk States Confinement as a Long Range Sensor for Impurities \\ and a Quantum Information Transfer Channel}

\author{O. Brovko}
\affiliation{Max-Planck-Institut f\"{u}r Mikrostrukturphysik, Weinberg 2, D06120 Halle, Germany}
\author{P. A. Ignatiev}
\affiliation{Max-Planck-Institut f\"{u}r Mikrostrukturphysik, Weinberg 2, D06120 Halle, Germany}
\author{V. S. Stepanyuk}
\affiliation{Max-Planck-Institut f\"{u}r Mikrostrukturphysik, Weinberg 2, D06120 Halle, Germany}

\date{23 November 2010}

\begin{abstract}
    We show that confinement of bulk electrons can be observed at low-dimensional surface structures and can serve as a long range sensor for the magnetism and electronic properties of single impurities or as a quantum information transfer channel with large coherence lengths. Our \emph{ab-initio} calculations reveal oscillations of electron density in magnetic chains on metallic surfaces and help to unambiguously identify the electrons involved as bulk electrons. We furthermore discuss a possibility to utilize bulk states confinement to transfer quantum information, encoded in an atom's species or spin, across distances of several nanometers with high efficiency.
\end{abstract}

\pacs{73.20.-r} \keywords{bulk states,chain,confinement}

\maketitle

% ---------------------------------------------------------------
% --------------  Intro  ----------------------------------------
% ---------------------------------------------------------------

    Doubtless one of the most popular research highlights of the past two decades has been the topic of quantum information and the inseparably linked questions of its storage and transportation. \cite{Bennett2000} This trend has lead not only to attempts of miniaturizing classical circuitry to the level of single-atom resolution \cite{Heinrich2002} but also to the birth of spintronics \cite{Bader2010}. The latter has, in turn, stimulated research in metal \cite{Chappert2008}, semiconductor \cite{Awschalom2007,*Dery2007}, organic \cite{Xiong2004,*Bogani2008} and nanostructured material \cite{Ardavan2007} systems aiming at utilizing atomic spins for quantum information storage. The transfer of quantum information (QIT), however, turned out to be a more challenging task. Since most prospective nanoscale information storage systems (molecules or magnetic nanostructures) are surface based, one needs an effective means of transporting information over the surface of a metal or semiconductor. In terms of a quantum system this means creating entanglement between the states of ``sending'' and ``receiving'' storage elements. In the case of spintronics one has to somehow couple the spins of single storage units. It has been shown that one way to achieve such entanglement (or plainly said - interaction) is to use the surface state (SS) \cite{Tamm1932,*Shockley1939} -- a state formed by electrons trapped between the repulsive vacuum potential at the surface and the projected bulk band gap. A major feature of those electrons is their free-electron-like behavior \cite{Shockley1939} in the plane of the surface. Localized potentials of surface impurities and structures can scatter SS electrons \cite{Friedel1958,*Crommie1993a} and even confine them to closed geometries \cite{Crommie1993,Manoharan2000,*Stepanyuk2005}. The large wave length of the SS and its weak coupling to the underlying bulk result in a large spatial coherence of SS electrons, allowing them to serve as mediators for the interatomic impurity interaction across distances of several nanometers. \cite{Abajo2010} For example, it has been shown that by carefully constructing a confining structure (an elliptic corral) one can use SS electrons to coherently project the electronic structure of a single adatom to a remote location on the surface \cite{Manoharan2000,*Stepanyuk2005}. However constructing complex artificial structures is a demanding task and the ability to support a surface state is not a commonplace feature among metallic and semiconductor surfaces. An obvious thing to do would be to switch from surface to bulk electrons for the mediation of interactions. Bulk states confinement is already very well known as the cause of interlayer exchange coupling and the giant magnetoresistance. \cite{Grunberg1986,*Baibich1988,*Bruno1991} Unfortunately bulk electrons have shorter coherence lengths due to the scattering at atomic cores and can propagate effectively only along certain directions determined by the geometry of the Fermi surface (see, f.e. Refs.~\onlinecite{Bruno1991} and \onlinecite{Weismann2009}). Quite recently experiments by Didiot and coworkers \cite{Didiot2010} have shown that bulk states can indeed exhibit electronic behavior very similar to the one observed for surface electrons. They have observed lateral confinement of bulk electrons in artificial and natural structures at Au(111) surfaces. \cite{Didiot2010} This newly gained realization brings with it an obvious question whether bulk states confinement at surfaces can be put to use in a field of quantum information transfer as it was hinted for surface states by Abajo~\emph{et al.} \cite{Abajo2010}

    The results of Didiot~\emph{et al.} \cite{Didiot2010} on the confinement of bulk electrons at surface structures have shown that such confinement is principally possible. However, to apply this concept to QIT one has find a way to precisely channel the confinement to allow directed information transfer. To achieve this goal one could turn to one of the most popular prototype systems in QIT and the most natural analogue of a classical wire -- a magnetic quantum spin chain \cite{Bose2003,*Christandl2004,*Gambardella2006}. In such a system an information quantum (a qubit) can be encoded in the spin state of a certain chain unit. Such chains attract continuous attention due to their rich quantum state entanglement features. The physical realization of the chain can be very different: coupled nanocavities \cite{Hartmann2007}, cold atoms in optical lattices \cite{Duan2003}, arrays of quantum dots, \emph{etc.} However, in most cases entanglement is accomplished by the local interaction of nearest neighbor spins \cite{Bose2003} which brings with itself a disadvantage of either a fast loss of fidelity with increasing chain length \cite{Bose2003} or of extremely complicated requirements to the coupling inhomogeneity. \cite{Eckert2007,*Murphy2010} In the present paper we show, that low dimensional surface structures, e.g. linear monoatomic metallic spin-chains on metal surfaces, can serve as a vessel for the directional confinement of bulk states, which in turn can provide the quantum entanglement of individual chain elements, necessary for the QIT, at distances in the nanometer range.

% ---------------------------------------------------------------
% --------------  methods  --------------------------------------
% ---------------------------------------------------------------

    As a tool for studying the confinement in low-dimensional surface structures we have used a code based on Korringa-Kohn-Rostoker (KKR) Green's function method in atomic spheres approximation. This method is described in detail in numerous publications \cite{Wildberger1995,*Zeller1995}. Here we will only mention that in our calculations we have made use of the local spin density and atomic spheres approximations. The KKR method has proven itself to be highly suitable for electron propagation and confinement calculations. \cite{Stepanyuk2006,Negulyaev2008}

% ---------------------------------------------------------------
% --------------  Results  --------------------------------------
% ---------------------------------------------------------------

    First of all, let us investigate, whether bulk states confinement can indeed be observed in low-dimensional surface structures. We consider metallic chains on a metallic surfaces as a model system (as a particular example we take a Co chain $35~\mathrm{\AA}$ in length on a Cu(111) surface). Such systems can be reliably constructed experimentally either by single-atom-manipulation \cite{Lagoute2007,Negulyaev2008} or by utilizing the self-assembly capabilities of e.g. stepped surfaces. \cite{Shiraki2004} Besides, monoatomic chains are known to exhibit collective spin excitation behavior \cite{Hirjibehedin2006} and can coherently channel excited electrons \cite{Diaz-Tendero2009prl,Diaz-Tendero2009} and enhance the interatomic interaction at large separations \cite{Brovko2008prl}. This is a good starting point for any investigation of quantum entanglement.

    \begin{figure*}
		\includegraphics[scale=1.0]{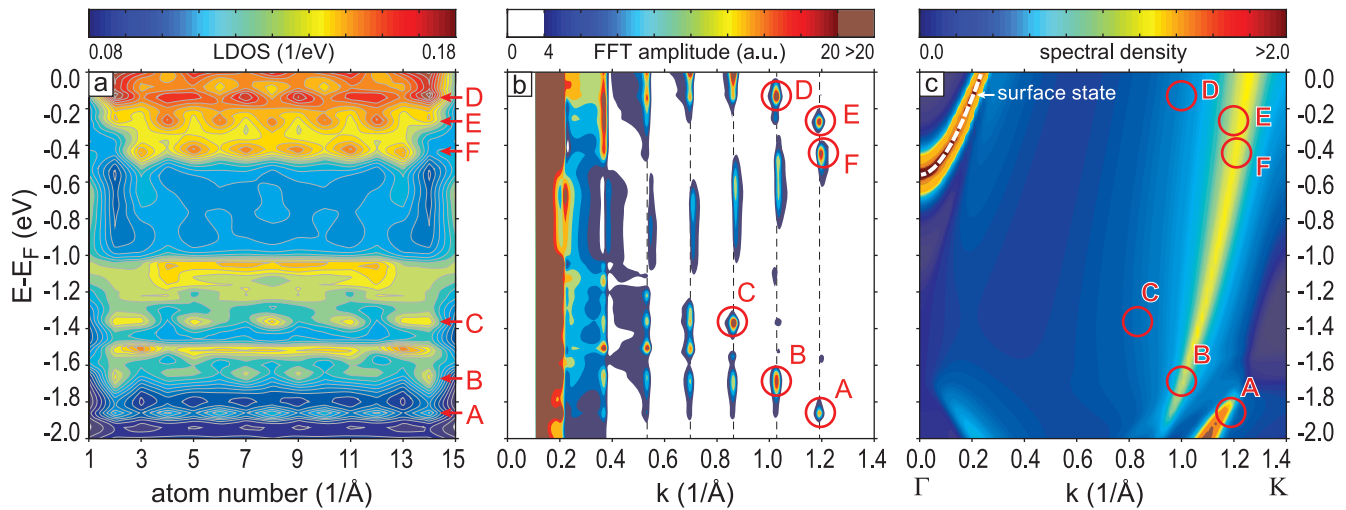}
        \caption{\label{fig:1}(color) a) Local density of $sp$ electronic states of each atom in the chain as a function of energy and the number of atom in the chain. To produce the map, the LDOS has been evaluated at each atomic sphere along the chain and then linearly interpolated in between the atomic positions to produce a smooth map for clearer presentation. Arrows and letters mark the most pronounced oscillations. b) 1D Fourier transform of the LDOS map. Circles and arrows mark the peaks corresponding to the oscillations marked in Fig.~\ref{fig:1}a by arrows. c) In-plane momentum resolved electron density map for the Cu(111) surface. Arrows and letters are marking the positions of peaks in Fig.~\ref{fig:1}b}
    \end{figure*}

    The easiest way to detect electron confinement is to study the spatially resolved electron density variation in the system. As we are talking of distances of the order of several nanometer we can safely assume that only $sp$ electrons are enough delocalized to actively participate in the confinement. In Fig.~\ref{fig:1} we plot the local density of $sp$ electronic states ($sp$-LDOS) of each atom in the chain as a function of energy and the number of atom in the chain. To produce the map, the LDOS has been evaluated at each atomic sphere along the chain and then linearly interpolated in between the atomic positions. The first thing that catches the eye is the presence of multiple regions where pronounced oscillation of the LDOS at a certain energy can be observed along the chain that cannot be expected to exist at either a clean surface or an infinite monatomic wire. This signifies that electrons responsible for the periodic density variation have coherence lengths exceeding the double length of the chain and that they are scattered and confined by the chain potential. We can witness quantum confinement of electrons at one-dimensional surface structures. Carefully studying the LDOS map in Fig.~\ref{fig:1} we can observe different oscillation modes of the chain. For example, at $-1.86$, $-1.68$ and $-1.38~\mathrm{eV}$ there are unmistakable oscillation with a period of about $5.3$, $6.1$ and $7.3~\mathrm{\AA}$ respectively (marked by A, B and C on the graphics). Similarly pronounced oscillations can be found at $-0.12$, $-0.28$ and $-0.43~\mathrm{eV}$ (D,E and F, respectively). This is, however, by far not a complete list of existing modes. There are doubtless modes with longer wavelengths present, yet the limitation of the chain length prevents them from being clearly discernable just by studying the atom resolved LDOS.

    To get a quantitative picture of the modes present in the chain we do a 1D Fourier transform of the LDOS map (Fig.~\ref{fig:1}b). The low-frequency part of the spectrum ($k < 0.4~\mathrm{\AA}^{-1}$) can not be considered meaningful, as the length of the chain is limited and the lowest wavevector we can reliably detect is about $k_{min} = \pi/L_{chain} \approx 0.1~\mathrm{\AA}^{-1}$, where $L_{chain}=38.4~\mathrm{\AA}$ is the effective length of the chain. Moreover, the edge effects in the chain (reduced electron density and edge states) manifest themselves in the spectrum as low-frequency inclusions, thus we will concentrate on the higher frequencies $k \in \sbr{0.4; 1.4}~\mathrm{\AA}^{-1}$ to analyze the electronic confinement. Indeed, starting from $0.37~\mathrm{\AA}^{-1}$ the background of the spectrum ebbs away and only pronounced equidistantly spaced peaks remain. The period of observed peaks, determined from the spacing of the peaks is $\Delta k=0.164\pm0.004~\mathrm{\AA}^{-1}$. In terms of a particle-in-a-box model this corresponds to a ``box-length'' of $L_{eff} = 38.3~\mathrm{\AA}$ which is almost exactly the extent of the chain mentioned above. The eigenfrequencies of a particle in such a box are marked in Fig.~\ref{fig:1}b by vertical dashed lines. Careful observation yields two important conclusions. For one, at each eigenfrequency confinement can be observed for electrons with certain energies, which reflects the underlying band structure of the system. However, while the spectrum maxima in the energy regions below $-1.0~\mathrm{eV}$ and above $-0.3~\mathrm{eV}$ match the eigenfrequencies extremely well, the peaks in the energy range between $-0.3$ and $-1.0~\mathrm{eV}$ seem to be slightly shifted to higher frequencies. To understand the physics behind this seeming discrepancy and at the same time determine the affiliation of the electrons participation in the confinement we take a look at the band structure of the surface on which our chain resides. A spectral density map (SDM) of $sp$ electrons with an in-plane $k$ vector ($k_{||}$) aligned along the $\Gamma-\mathrm{K}$ direction of a Cu(111) surface is presented in Fig.~\ref{fig:1}c.

    The question of the origin of electrons participating in the confinement is quite easily solved if we consider that the Cu surface state is represented by a parabolic band starting from $-0.45~\mathrm{eV}$ upwards for Cu(111) and with $k_{||}$ momenta not exceeding the Fermi wave vector $k_{F}^{Cu}\approx0.23~\mathrm{\AA}^{-1}$ (upper left corner of Fig.~\ref{fig:1}c). The remaining part of the SDM reflects a projection of a Cu bulk band structure onto a (111) surface of the crystal. If one compares the band structure of the surface with the Fourier spectrum of the oscillations we observe in the Co chain it immediately becomes clear that most of the confinement must be attributed to either the bulk states resonance scattering and confinement in surface structures or to the confinement of the chain's own $sp$ electrons. To hold apart the latter two we can compare the maxima of the space resolved LDOS spectrum in Fig.~\ref{fig:1}b with the SDM in Fig.~\ref{fig:1}c. As an example, let us take the most pronounced oscillations we could identify in the atom resolved LDOS map (Fig.~\ref{fig:1}a). The spectral maxima corresponding to those oscillations are marked by red circles and corresponding letters in Fig.~\ref{fig:1}b. If we now project those maxima onto the spectral density map, we can conclude that the maxima A and B clearly coincide with the maximum intensity areas of a surface projected band of $p$ character at the wave vector predefined by the eigenmode of the chain. They can be thus attributed to the bulk states confinement and resonance scattering. The same cannot be said about the maxima C and D as they seem to lie in regions where the SDM is mostly featureless and cannot account for the energy selection. Those peaks can thus be attributed to the confinement of the chain's own $sp$-electrons or the electrons selectively scattered into the chain due to the coupling of the chain, as a 1D quantum system to the bulk states continuum. The peaks E and F are a continuation of the riddle of shifted peak positions. As is seen from Fig.~\ref{fig:1}b they both seem to belong to the mode with a wave vector of $\sim1.19~\mathrm{\AA}^{-1}$, yet the peak at $-0.45~\mathrm{eV}$ is unmistakably shifted towards higher frequencies by about $0.01~\mathrm{\AA}^{-1}$. The origin of that is to be sought in the kind of electrons constituting the peaks. Our calculation show, that region of the SDM contained between $-0.3$ and $-1.0~\mathrm{eV}$ is dominated by the $s$ band of the Cu(111) surface, while the remaining part of the SDM is mostly composed by $p$ electrons. Since $s$ and $p$ electrons obtain different scattering phases at the potential boundary, they also have different confinement lengths and are subject to different eigenmodes. Another confirmation for that can be found in the fact, that in Fig.~\ref{fig:1}a oscillations marked E and F display virtually the same periodicity but have phases differing by $\pi/2$.

    Up to now we have not made any differentiation between electrons of different spin. Yet the LDOS of the chain is highly polarized. The majority band is centered around $-1.5~\mathrm{eV}$ and extends up to $-1.0~\mathrm{eV}$. The minority band is located close to the Fermi energy and reaches down to energies of about $-0.8~\mathrm{eV}$. Those regions exactly coincide with the regions of most effective confinement. Indeed, analysis of spin-polarization shows that oscillations seen in Fig.~\ref{fig:1}a below $-1.0~\mathrm{eV}$ are predominantly of majority character while the upper energy range ($E-E_F>-0.8~\mathrm{eV}$) is mainly minority.

    \begin{figure}
		\includegraphics{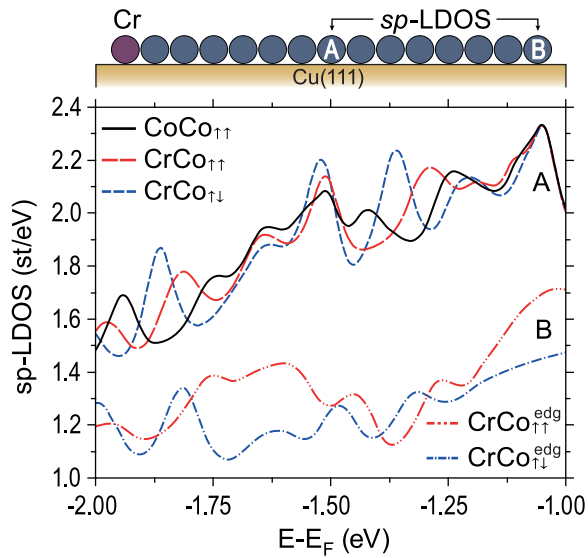}
        \caption{\label{fig:2}(color online) The local density of $sp$-electrons at the central chain atom for a ferromagnetic Co chain (solid black curves), a Co chain with the edge atom replaces by Cr, ferromagnetically (red long dash) or antiferromagnetically (blue short dash) coupled to the rest of the chain. The local density of $p$-electrons at the edge chain atom for a Co chain with the opposite-edge atom replaces by Cr, ferromagnetically (red dash-dot-dot) or antiferromagnetically (blue short dash-dot) coupled to the rest of the chain.}
    \end{figure}

    The properties of a chain described above are a fair starting point for possible applications in the field of quantum information storage and transfer. In the remaining part of the Letter we would like to discuss a possible example of such an application -- transfer of magnetically encoded information using the confinement of bulk states in a 1D Co chain on a Cu(111) surface. To avoid misunderstanding we would like to mention from the start, that we use the term information transfer in the spirit of the works by Manoharan and coworkers \cite{Manoharan2000}. The information transfer scheme we propose is quite simple. The information can be encoded in the spin orientation of the edge atom and is then read out at a certain point of the chain. To achieve that we will use the fact that confinement can be manipulated by changing the boundary conditions for the confined electrons. In case of a one-dimensional chain this would mean changing the potential terminating the chain. This can be achieved in several ways. One way would be to change the atom terminating the chain. Let us focus first on the central atom of the chain (thus eliminating the possibility of interference of edge effects) and see how the its $sp$ LDOS would change if we replace one of the edge Co atoms with, say, chromium (see the sketch at the top of Fig.~\ref{fig:2}). The $sp$-LDOS of the middle Co atom in a homogeneous Co chain is shown in Fig.~\ref{fig:2} by a solid black curve. If we now replace one of the edge Co atoms by a Cr, coupled ferromagnetically to the rest of the chain, we would find that the LDOS of the central Co adatom has changed and now looks like as shown by the red long-dashed curve. One can observe clear peak shifts caused by the change in the boundary conditions which means, that the LDOS of the central atom of a an over $3~\mathrm{nm}$ long chain is sensitive to the changes introduced at the chain's edges. Furthermore, knowing that the confinement is spin-sensitive, we can now reverse the spin direction of the Cr and look for further changes in the confinement. The LDOS of the central adatom for such a case (when Cr is coupled antiferromagnetically to the rest of the chain) is presented by the blue short-dashed curve in Fig.~\ref{fig:2}. The peaks have once again shifted their positions indicating, that 1D confinement in a magnetic chain is indeed very much spin sensitive.

    To verify that the effects described above are not unique to the central atom of the chain, we present in Fig.~\ref{fig:2} two additional curves (red dash-dot-dotted and blue dash-dotted) representing the $sp$-LDOS of the edge Co atom in a 15-atomic chain terminated at the opposite end by a Cr coupled ferromagnetically (respectively antiferromagnetically) to the rest of the chain. One can see that even here, the peak shifts caused by a spin-flip at the other end of the chain are clearly discernable. This proves that the information about the spin state of the edge atom can be transferred to a remote location coherently and with high fidelity.

    We can thus summarize that bulk states confinement has been shown to exist in one-dimensional mono-atomic metallic spin-chains on metallic surfaces. It manifest itself as standing waves of the electron density with wave vectors predefined by the eigenmodes of the chain and can be sensed at all points of the chain. It is, moreover, highly sensitive to changes in boundary conditions which allows one to register the change in electronic and magnetic properties of single atoms of the chain across distances of at least several nanometers. We believe that observing this phenomenon experimentally is quite feasible with modern state-of-the-art experimental techniques (STM, AFM) and may prove to be an interesting and rewarding challenge.

    %\bibliography{booBib}

\begin{thebibliography}{39}%
\makeatletter
\providecommand \@ifxundefined [1]{%
 \@ifx{#1\undefined}
}%
\providecommand \@ifnum [1]{%
 \ifnum #1\expandafter \@firstoftwo
 \else \expandafter \@secondoftwo
 \fi
}%
\providecommand \@ifx [1]{%
 \ifx #1\expandafter \@firstoftwo
 \else \expandafter \@secondoftwo
 \fi
}%
\providecommand \natexlab [1]{#1}%
\providecommand \enquote  [1]{``#1''}%
\providecommand \bibnamefont  [1]{#1}%
\providecommand \bibfnamefont [1]{#1}%
\providecommand \citenamefont [1]{#1}%
\providecommand \href@noop [0]{\@secondoftwo}%
\providecommand \href [0]{\begingroup \@sanitize@url \@href}%
\providecommand \@href[1]{\@@startlink{#1}\@@href}%
\providecommand \@@href[1]{\endgroup#1\@@endlink}%
\providecommand \@sanitize@url [0]{\catcode `\\12\catcode `\$12\catcode
  `\&12\catcode `\#12\catcode `\^12\catcode `\_12\catcode `\%12\relax}%
\providecommand \@@startlink[1]{}%
\providecommand \@@endlink[0]{}%
\providecommand \url  [0]{\begingroup\@sanitize@url \@url }%
\providecommand \@url [1]{\endgroup\@href {#1}{\urlprefix }}%
\providecommand \urlprefix  [0]{URL }%
\providecommand \Eprint [0]{\href }%
\providecommand \doibase [0]{http://dx.doi.org/}%
\providecommand \selectlanguage [0]{\@gobble}%
\providecommand \bibinfo  [0]{\@secondoftwo}%
\providecommand \bibfield  [0]{\@secondoftwo}%
\providecommand \translation [1]{[#1]}%
\providecommand \BibitemOpen [0]{}%
\providecommand \bibitemStop [0]{}%
\providecommand \bibitemNoStop [0]{.\EOS\space}%
\providecommand \EOS [0]{\spacefactor3000\relax}%
\providecommand \BibitemShut  [1]{\csname bibitem#1\endcsname}%
\let\auto@bib@innerbib\@empty
%</preamble>
\bibitem [{\citenamefont {Bennett}\ and\ \citenamefont
  {DiVincenzo}(2000)}]{Bennett2000}%
  \BibitemOpen
  \bibfield  {author} {\bibinfo {author} {\bibfnamefont {C.~H.}\ \bibnamefont
  {Bennett}}\ and\ \bibinfo {author} {\bibfnamefont {D.~P.}\ \bibnamefont
  {DiVincenzo}},\ }\href {http://dx.doi.org/10.1038/35005001} {\bibfield
  {journal} {\bibinfo  {journal} {Nature}\ }\textbf {\bibinfo {volume} {404}},\
  \bibinfo {pages} {247} (\bibinfo {year} {2000})}\BibitemShut {NoStop}%
\bibitem [{\citenamefont {Heinrich}\ \emph {et~al.}(2002)\citenamefont
  {Heinrich}, \citenamefont {Lutz}, \citenamefont {Gupta},\ and\ \citenamefont
  {Eigler}}]{Heinrich2002}%
  \BibitemOpen
  \bibfield  {author} {\bibinfo {author} {\bibfnamefont {A.~J.}\ \bibnamefont
  {Heinrich}}, \bibinfo {author} {\bibfnamefont {C.~P.}\ \bibnamefont {Lutz}},
  \bibinfo {author} {\bibfnamefont {J.~A.}\ \bibnamefont {Gupta}}, \ and\
  \bibinfo {author} {\bibfnamefont {D.~M.}\ \bibnamefont {Eigler}},\ }\href
  {http://www.jstor.org/stable/3832783} {\bibfield  {journal} {\bibinfo
  {journal} {Science}\ }\bibinfo {series} {New Series},\ \textbf {\bibinfo
  {volume} {298}},\ \bibinfo {pages} {pp. 1381} (\bibinfo {year}
  {2002})}\BibitemShut {NoStop}%
\bibitem [{\citenamefont {Bader}\ and\ \citenamefont
  {Parkin}(2010)}]{Bader2010}%
  \BibitemOpen
  \bibfield  {author} {\bibinfo {author} {\bibfnamefont {S.}~\bibnamefont
  {Bader}}\ and\ \bibinfo {author} {\bibfnamefont {S.}~\bibnamefont {Parkin}},\
  }\href {\doibase 10.1146/annurev-conmatphys-070909-104123} {\bibfield
  {journal} {\bibinfo  {journal} {Ann. Rev. Cond. Matter Phys.}\ }\textbf
  {\bibinfo {volume} {1}},\ \bibinfo {pages} {71} (\bibinfo {year}
  {2010})}\BibitemShut {NoStop}%
\bibitem [{\citenamefont {Chappert}\ and\ \citenamefont
  {Kim}(2008)}]{Chappert2008}%
  \BibitemOpen
  \bibfield  {author} {\bibinfo {author} {\bibfnamefont {C.}~\bibnamefont
  {Chappert}}\ and\ \bibinfo {author} {\bibfnamefont {J.-V.}\ \bibnamefont
  {Kim}},\ }\href {http://dx.doi.org/10.1038/nphys1122} {\bibfield  {journal}
  {\bibinfo  {journal} {Nat Phys}\ }\textbf {\bibinfo {volume} {4}},\ \bibinfo
  {pages} {837} (\bibinfo {year} {2008})}\BibitemShut {NoStop}%
\bibitem [{\citenamefont {Awschalom}\ and\ \citenamefont
  {Flatte}(2007)}]{Awschalom2007}%
  \BibitemOpen
  \bibfield  {author} {\bibinfo {author} {\bibfnamefont {D.~D.}\ \bibnamefont
  {Awschalom}}\ and\ \bibinfo {author} {\bibfnamefont {M.~E.}\ \bibnamefont
  {Flatte}},\ }\href {\doibase 10.1038/nphys551} {\bibfield  {journal}
  {\bibinfo  {journal} {Nat Phys}\ }\textbf {\bibinfo {volume} {3}},\ \bibinfo
  {pages} {153} (\bibinfo {year} {2007})}\BibitemShut {NoStop}%
\bibitem [{\citenamefont {Dery}\ \emph {et~al.}(2007)\citenamefont {Dery},
  \citenamefont {Dalal}, \citenamefont {Cywinski},\ and\ \citenamefont
  {Sham}}]{Dery2007}%
  \BibitemOpen
  \bibfield  {author} {\bibinfo {author} {\bibfnamefont {H.}~\bibnamefont
  {Dery}}, \bibinfo {author} {\bibfnamefont {P.}~\bibnamefont {Dalal}},
  \bibinfo {author} {\bibfnamefont {L.}~\bibnamefont {Cywinski}}, \ and\
  \bibinfo {author} {\bibfnamefont {L.~J.}\ \bibnamefont {Sham}},\ }\href
  {http://dx.doi.org/10.1038/nature05833} {\bibfield  {journal} {\bibinfo
  {journal} {Nature}\ }\textbf {\bibinfo {volume} {447}},\ \bibinfo {pages}
  {573} (\bibinfo {year} {2007})}\BibitemShut {NoStop}%
\bibitem [{\citenamefont {Xiong}\ \emph {et~al.}(2004)\citenamefont {Xiong},
  \citenamefont {Wu}, \citenamefont {Valy~Vardeny},\ and\ \citenamefont
  {Shi}}]{Xiong2004}%
  \BibitemOpen
  \bibfield  {author} {\bibinfo {author} {\bibfnamefont {Z.~H.}\ \bibnamefont
  {Xiong}}, \bibinfo {author} {\bibfnamefont {D.}~\bibnamefont {Wu}}, \bibinfo
  {author} {\bibfnamefont {Z.}~\bibnamefont {Valy~Vardeny}}, \ and\ \bibinfo
  {author} {\bibfnamefont {J.}~\bibnamefont {Shi}},\ }\href
  {http://dx.doi.org/10.1038/nature02325} {\bibfield  {journal} {\bibinfo
  {journal} {Nature}\ }\textbf {\bibinfo {volume} {427}},\ \bibinfo {pages}
  {821} (\bibinfo {year} {2004})}\BibitemShut {NoStop}%
\bibitem [{\citenamefont {Bogani}\ and\ \citenamefont
  {Wernsdorfer}(2008)}]{Bogani2008}%
  \BibitemOpen
  \bibfield  {author} {\bibinfo {author} {\bibfnamefont {L.}~\bibnamefont
  {Bogani}}\ and\ \bibinfo {author} {\bibfnamefont {W.}~\bibnamefont
  {Wernsdorfer}},\ }\href {http://dx.doi.org/10.1038/nmat2133} {\bibfield
  {journal} {\bibinfo  {journal} {Nat Mater}\ }\textbf {\bibinfo {volume}
  {7}},\ \bibinfo {pages} {179} (\bibinfo {year} {2008})}\BibitemShut {NoStop}%
\bibitem [{\citenamefont {Ardavan}\ \emph {et~al.}(2007)\citenamefont
  {Ardavan}, \citenamefont {Rival}, \citenamefont {Morton}, \citenamefont
  {Blundell}, \citenamefont {Tyryshkin}, \citenamefont {Timco},\ and\
  \citenamefont {Winpenny}}]{Ardavan2007}%
  \BibitemOpen
  \bibfield  {author} {\bibinfo {author} {\bibfnamefont {A.}~\bibnamefont
  {Ardavan}}, \bibinfo {author} {\bibfnamefont {O.}~\bibnamefont {Rival}},
  \bibinfo {author} {\bibfnamefont {J.~J.~L.}\ \bibnamefont {Morton}}, \bibinfo
  {author} {\bibfnamefont {S.~J.}\ \bibnamefont {Blundell}}, \bibinfo {author}
  {\bibfnamefont {A.~M.}\ \bibnamefont {Tyryshkin}}, \bibinfo {author}
  {\bibfnamefont {G.~A.}\ \bibnamefont {Timco}}, \ and\ \bibinfo {author}
  {\bibfnamefont {R.~E.~P.}\ \bibnamefont {Winpenny}},\ }\href {\doibase
  10.1103/PhysRevLett.98.057201} {\bibfield  {journal} {\bibinfo  {journal}
  {Phys. Rev. Lett.}\ }\textbf {\bibinfo {volume} {98}},\ \bibinfo {pages}
  {057201} (\bibinfo {year} {2007})}\BibitemShut {NoStop}%
\bibitem [{\citenamefont {Tamm}(1932)}]{Tamm1932}%
  \BibitemOpen
  \bibfield  {author} {\bibinfo {author} {\bibfnamefont {I.}~\bibnamefont
  {Tamm}},\ }\href {\doibase 10.1007/BF01341581} {\bibfield  {journal}
  {\bibinfo  {journal} {Zeitschrift fur Physik}\ }\textbf {\bibinfo {volume}
  {76}},\ \bibinfo {pages} {849} (\bibinfo {year} {1932})}\BibitemShut
  {NoStop}%
\bibitem [{\citenamefont {Shockley}(1939)}]{Shockley1939}%
  \BibitemOpen
  \bibfield  {author} {\bibinfo {author} {\bibfnamefont {W.}~\bibnamefont
  {Shockley}},\ }\href {\doibase 10.1103/PhysRev.56.317} {\bibfield  {journal}
  {\bibinfo  {journal} {Phys. Rev.}\ }\textbf {\bibinfo {volume} {56}},\
  \bibinfo {pages} {317} (\bibinfo {year} {1939})}\BibitemShut {NoStop}%
\bibitem [{\citenamefont {Friedel}(1958)}]{Friedel1958}%
  \BibitemOpen
  \bibfield  {author} {\bibinfo {author} {\bibfnamefont {J.}~\bibnamefont
  {Friedel}},\ }\href@noop {} {\bibfield  {journal} {\bibinfo  {journal} {Nuovo
  Cimento}\ }\textbf {\bibinfo {volume} {7}},\ \bibinfo {pages} {287} (\bibinfo
  {year} {1958})}\BibitemShut {NoStop}%
\bibitem [{\citenamefont {Crommie}\ \emph
  {et~al.}(1993{\natexlab{a}})\citenamefont {Crommie}, \citenamefont {Lutz},\
  and\ \citenamefont {Eigler}}]{Crommie1993a}%
  \BibitemOpen
  \bibfield  {author} {\bibinfo {author} {\bibfnamefont {M.~F.}\ \bibnamefont
  {Crommie}}, \bibinfo {author} {\bibfnamefont {C.~P.}\ \bibnamefont {Lutz}}, \
  and\ \bibinfo {author} {\bibfnamefont {D.~M.}\ \bibnamefont {Eigler}},\
  }\href@noop {} {\bibfield  {journal} {\bibinfo  {journal} {Nature}\ }\textbf
  {\bibinfo {volume} {363}},\ \bibinfo {pages} {524} (\bibinfo {year}
  {1993}{\natexlab{a}})}\BibitemShut {NoStop}%
\bibitem [{\citenamefont {Crommie}\ \emph
  {et~al.}(1993{\natexlab{b}})\citenamefont {Crommie}, \citenamefont {Lutz},\
  and\ \citenamefont {Eigler}}]{Crommie1993}%
  \BibitemOpen
  \bibfield  {author} {\bibinfo {author} {\bibfnamefont {M.~F.}\ \bibnamefont
  {Crommie}}, \bibinfo {author} {\bibfnamefont {C.~P.}\ \bibnamefont {Lutz}}, \
  and\ \bibinfo {author} {\bibfnamefont {D.~M.}\ \bibnamefont {Eigler}},\
  }\href {\doibase 10.1126/science.262.5131.218} {\bibfield  {journal}
  {\bibinfo  {journal} {Science}\ }\textbf {\bibinfo {volume} {262}},\ \bibinfo
  {pages} {218} (\bibinfo {year} {1993}{\natexlab{b}})}\BibitemShut {NoStop}%
\bibitem [{\citenamefont {Manoharan}\ \emph {et~al.}(2000)\citenamefont
  {Manoharan}, \citenamefont {Lutz},\ and\ \citenamefont
  {Eigler}}]{Manoharan2000}%
  \BibitemOpen
  \bibfield  {author} {\bibinfo {author} {\bibfnamefont {H.~C.}\ \bibnamefont
  {Manoharan}}, \bibinfo {author} {\bibfnamefont {C.~P.}\ \bibnamefont {Lutz}},
  \ and\ \bibinfo {author} {\bibfnamefont {D.~M.}\ \bibnamefont {Eigler}},\
  }\href@noop {} {\bibfield  {journal} {\bibinfo  {journal} {Nature}\ }\textbf
  {\bibinfo {volume} {403}},\ \bibinfo {pages} {512} (\bibinfo {year}
  {2000})}\BibitemShut {NoStop}%
\bibitem [{\citenamefont {Stepanyuk}\ \emph {et~al.}(2005)\citenamefont
  {Stepanyuk}, \citenamefont {Niebergall}, \citenamefont {Hergert},\ and\
  \citenamefont {Bruno}}]{Stepanyuk2005}%
  \BibitemOpen
  \bibfield  {author} {\bibinfo {author} {\bibfnamefont {V.~S.}\ \bibnamefont
  {Stepanyuk}}, \bibinfo {author} {\bibfnamefont {L.}~\bibnamefont
  {Niebergall}}, \bibinfo {author} {\bibfnamefont {W.}~\bibnamefont {Hergert}},
  \ and\ \bibinfo {author} {\bibfnamefont {P.}~\bibnamefont {Bruno}},\ }\href
  {\doibase 10.1103/PhysRevLett.94.187201} {\bibfield  {journal} {\bibinfo
  {journal} {Phys. Rev. Lett.}\ }\textbf {\bibinfo {volume} {94}},\ \bibinfo
  {eid} {187201} (\bibinfo {year} {2005})}\BibitemShut {NoStop}%
\bibitem [{\citenamefont {de~Abajo}\ \emph {et~al.}(2010)\citenamefont
  {de~Abajo}, \citenamefont {Cordon}, \citenamefont {Corso}, \citenamefont
  {Schiller},\ and\ \citenamefont {Ortega}}]{Abajo2010}%
  \BibitemOpen
  \bibfield  {author} {\bibinfo {author} {\bibfnamefont {F.~J.~G.}\
  \bibnamefont {de~Abajo}}, \bibinfo {author} {\bibfnamefont {J.}~\bibnamefont
  {Cordon}}, \bibinfo {author} {\bibfnamefont {M.}~\bibnamefont {Corso}},
  \bibinfo {author} {\bibfnamefont {F.}~\bibnamefont {Schiller}}, \ and\
  \bibinfo {author} {\bibfnamefont {J.~E.}\ \bibnamefont {Ortega}},\ }\href
  {\doibase 10.1039/b9nr00386j} {\bibfield  {journal} {\bibinfo  {journal}
  {Nanoscale}\ }\textbf {\bibinfo {volume} {2}},\ \bibinfo {pages} {717}
  (\bibinfo {year} {2010})}\BibitemShut {NoStop}%
\bibitem [{\citenamefont {Gr{\"u}nberg}\ \emph {et~al.}(1986)\citenamefont
  {Gr{\"u}nberg}, \citenamefont {Schreiber}, \citenamefont {Pang},
  \citenamefont {Brodsky},\ and\ \citenamefont {Sowers}}]{Grunberg1986}%
  \BibitemOpen
  \bibfield  {author} {\bibinfo {author} {\bibfnamefont {P.}~\bibnamefont
  {Gr{\"u}nberg}}, \bibinfo {author} {\bibfnamefont {R.}~\bibnamefont
  {Schreiber}}, \bibinfo {author} {\bibfnamefont {Y.}~\bibnamefont {Pang}},
  \bibinfo {author} {\bibfnamefont {M.~B.}\ \bibnamefont {Brodsky}}, \ and\
  \bibinfo {author} {\bibfnamefont {H.}~\bibnamefont {Sowers}},\ }\href
  {\doibase 10.1103/PhysRevLett.57.2442} {\bibfield  {journal} {\bibinfo
  {journal} {Phys. Rev. Lett.}\ }\textbf {\bibinfo {volume} {57}},\ \bibinfo
  {pages} {2442} (\bibinfo {year} {1986})}\BibitemShut {NoStop}%
\bibitem [{\citenamefont {Baibich}\ \emph {et~al.}(1988)\citenamefont
  {Baibich}, \citenamefont {Broto}, \citenamefont {Fert}, \citenamefont {{Van
  Dau}}, \citenamefont {Petroff}, \citenamefont {Eitenne}, \citenamefont
  {Creuzet}, \citenamefont {Friederich},\ and\ \citenamefont
  {Chazelas}}]{Baibich1988}%
  \BibitemOpen
  \bibfield  {author} {\bibinfo {author} {\bibfnamefont {M.~N.}\ \bibnamefont
  {Baibich}}, \bibinfo {author} {\bibfnamefont {J.~M.}\ \bibnamefont {Broto}},
  \bibinfo {author} {\bibfnamefont {A.}~\bibnamefont {Fert}}, \bibinfo {author}
  {\bibfnamefont {F.~N.}\ \bibnamefont {{Van Dau}}}, \bibinfo {author}
  {\bibfnamefont {F.}~\bibnamefont {Petroff}}, \bibinfo {author} {\bibfnamefont
  {P.}~\bibnamefont {Eitenne}}, \bibinfo {author} {\bibfnamefont
  {G.}~\bibnamefont {Creuzet}}, \bibinfo {author} {\bibfnamefont
  {A.}~\bibnamefont {Friederich}}, \ and\ \bibinfo {author} {\bibfnamefont
  {J.}~\bibnamefont {Chazelas}},\ }\href {\doibase 10.1103/PhysRevLett.61.2472}
  {\bibfield  {journal} {\bibinfo  {journal} {Phys. Rev. Lett.}\ }\textbf
  {\bibinfo {volume} {61}},\ \bibinfo {pages} {2472} (\bibinfo {year}
  {1988})}\BibitemShut {NoStop}%
\bibitem [{\citenamefont {Bruno}\ and\ \citenamefont
  {Chappert}(1991)}]{Bruno1991}%
  \BibitemOpen
  \bibfield  {author} {\bibinfo {author} {\bibfnamefont {P.}~\bibnamefont
  {Bruno}}\ and\ \bibinfo {author} {\bibfnamefont {C.}~\bibnamefont
  {Chappert}},\ }\href {\doibase 10.1103/PhysRevLett.67.1602} {\bibfield
  {journal} {\bibinfo  {journal} {Phys. Rev. Lett.}\ }\textbf {\bibinfo
  {volume} {67}},\ \bibinfo {pages} {1602} (\bibinfo {year}
  {1991})}\BibitemShut {NoStop}%
\bibitem [{\citenamefont {Weismann}\ \emph {et~al.}(2009)\citenamefont
  {Weismann}, \citenamefont {Wenderoth}, \citenamefont {Lounis}, \citenamefont
  {Zahn}, \citenamefont {Quaas}, \citenamefont {Ulbrich}, \citenamefont
  {Dederichs},\ and\ \citenamefont {Blugel}}]{Weismann2009}%
  \BibitemOpen
  \bibfield  {author} {\bibinfo {author} {\bibfnamefont {A.}~\bibnamefont
  {Weismann}}, \bibinfo {author} {\bibfnamefont {M.}~\bibnamefont {Wenderoth}},
  \bibinfo {author} {\bibfnamefont {S.}~\bibnamefont {Lounis}}, \bibinfo
  {author} {\bibfnamefont {P.}~\bibnamefont {Zahn}}, \bibinfo {author}
  {\bibfnamefont {N.}~\bibnamefont {Quaas}}, \bibinfo {author} {\bibfnamefont
  {R.~G.}\ \bibnamefont {Ulbrich}}, \bibinfo {author} {\bibfnamefont {P.~H.}\
  \bibnamefont {Dederichs}}, \ and\ \bibinfo {author} {\bibfnamefont
  {S.}~\bibnamefont {Blugel}},\ }\href {\doibase 10.1126/science.1168738}
  {\bibfield  {journal} {\bibinfo  {journal} {Science}\ }\textbf {\bibinfo
  {volume} {323}},\ \bibinfo {pages} {1190} (\bibinfo {year}
  {2009})}\BibitemShut {NoStop}%
\bibitem [{\citenamefont {Didiot}\ \emph {et~al.}(2010)\citenamefont {Didiot},
  \citenamefont {Cherkez}, \citenamefont {Kierren}, \citenamefont
  {Fagot-Revurat},\ and\ \citenamefont {Malterre}}]{Didiot2010}%
  \BibitemOpen
  \bibfield  {author} {\bibinfo {author} {\bibfnamefont {C.}~\bibnamefont
  {Didiot}}, \bibinfo {author} {\bibfnamefont {V.}~\bibnamefont {Cherkez}},
  \bibinfo {author} {\bibfnamefont {B.}~\bibnamefont {Kierren}}, \bibinfo
  {author} {\bibfnamefont {Y.}~\bibnamefont {Fagot-Revurat}}, \ and\ \bibinfo
  {author} {\bibfnamefont {D.}~\bibnamefont {Malterre}},\ }\href {\doibase
  10.1103/PhysRevB.81.075421} {\bibfield  {journal} {\bibinfo  {journal} {Phys.
  Rev. B}\ }\textbf {\bibinfo {volume} {81}},\ \bibinfo {pages} {075421}
  (\bibinfo {year} {2010})}\BibitemShut {NoStop}%
\bibitem [{\citenamefont {Bose}(2003)}]{Bose2003}%
  \BibitemOpen
  \bibfield  {author} {\bibinfo {author} {\bibfnamefont {S.}~\bibnamefont
  {Bose}},\ }\href {\doibase 10.1103/PhysRevLett.91.207901} {\bibfield
  {journal} {\bibinfo  {journal} {Phys. Rev. Lett.}\ }\textbf {\bibinfo
  {volume} {91}},\ \bibinfo {pages} {207901} (\bibinfo {year}
  {2003})}\BibitemShut {NoStop}%
\bibitem [{\citenamefont {Christandl}\ \emph {et~al.}(2004)\citenamefont
  {Christandl}, \citenamefont {Datta}, \citenamefont {Ekert},\ and\
  \citenamefont {Landahl}}]{Christandl2004}%
  \BibitemOpen
  \bibfield  {author} {\bibinfo {author} {\bibfnamefont {M.}~\bibnamefont
  {Christandl}}, \bibinfo {author} {\bibfnamefont {N.}~\bibnamefont {Datta}},
  \bibinfo {author} {\bibfnamefont {A.}~\bibnamefont {Ekert}}, \ and\ \bibinfo
  {author} {\bibfnamefont {A.~J.}\ \bibnamefont {Landahl}},\ }\href {\doibase
  10.1103/PhysRevLett.92.187902} {\bibfield  {journal} {\bibinfo  {journal}
  {Phys. Rev. Lett.}\ }\textbf {\bibinfo {volume} {92}},\ \bibinfo {pages}
  {187902} (\bibinfo {year} {2004})}\BibitemShut {NoStop}%
\bibitem [{\citenamefont {Gambardella}(2006)}]{Gambardella2006}%
  \BibitemOpen
  \bibfield  {author} {\bibinfo {author} {\bibfnamefont {P.}~\bibnamefont
  {Gambardella}},\ }\href {http://dx.doi.org/10.1038/nmat1662} {\bibfield
  {journal} {\bibinfo  {journal} {Nat Mater}\ }\textbf {\bibinfo {volume}
  {5}},\ \bibinfo {pages} {431} (\bibinfo {year} {2006})}\BibitemShut {NoStop}%
\bibitem [{\citenamefont {Hartmann}\ \emph {et~al.}(2007)\citenamefont
  {Hartmann}, \citenamefont {Brand\~ao},\ and\ \citenamefont
  {Plenio}}]{Hartmann2007}%
  \BibitemOpen
  \bibfield  {author} {\bibinfo {author} {\bibfnamefont {M.~J.}\ \bibnamefont
  {Hartmann}}, \bibinfo {author} {\bibfnamefont {F.~G. S.~L.}\ \bibnamefont
  {Brand\~ao}}, \ and\ \bibinfo {author} {\bibfnamefont {M.~B.}\ \bibnamefont
  {Plenio}},\ }\href {\doibase 10.1103/PhysRevLett.99.160501} {\bibfield
  {journal} {\bibinfo  {journal} {Phys. Rev. Lett.}\ }\textbf {\bibinfo
  {volume} {99}},\ \bibinfo {pages} {160501} (\bibinfo {year}
  {2007})}\BibitemShut {NoStop}%
\bibitem [{\citenamefont {Duan}\ \emph {et~al.}(2003)\citenamefont {Duan},
  \citenamefont {Demler},\ and\ \citenamefont {Lukin}}]{Duan2003}%
  \BibitemOpen
  \bibfield  {author} {\bibinfo {author} {\bibfnamefont {L.-M.}\ \bibnamefont
  {Duan}}, \bibinfo {author} {\bibfnamefont {E.}~\bibnamefont {Demler}}, \ and\
  \bibinfo {author} {\bibfnamefont {M.~D.}\ \bibnamefont {Lukin}},\ }\href
  {\doibase 10.1103/PhysRevLett.91.090402} {\bibfield  {journal} {\bibinfo
  {journal} {Phys. Rev. Lett.}\ }\textbf {\bibinfo {volume} {91}},\ \bibinfo
  {pages} {090402} (\bibinfo {year} {2003})}\BibitemShut {NoStop}%
\bibitem [{\citenamefont {Eckert}\ \emph {et~al.}(2007)\citenamefont {Eckert},
  \citenamefont {Romero-Isart},\ and\ \citenamefont {Sanpera}}]{Eckert2007}%
  \BibitemOpen
  \bibfield  {author} {\bibinfo {author} {\bibfnamefont {K.}~\bibnamefont
  {Eckert}}, \bibinfo {author} {\bibfnamefont {O.}~\bibnamefont
  {Romero-Isart}}, \ and\ \bibinfo {author} {\bibfnamefont {A.}~\bibnamefont
  {Sanpera}},\ }\href {http://stacks.iop.org/1367-2630/9/i=5/a=155} {\bibfield
  {journal} {\bibinfo  {journal} {New J. Phys.}\ }\textbf {\bibinfo {volume}
  {9}},\ \bibinfo {pages} {155} (\bibinfo {year} {2007})}\BibitemShut {NoStop}%
\bibitem [{\citenamefont {Murphy}\ \emph {et~al.}(2010)\citenamefont {Murphy},
  \citenamefont {Montangero}, \citenamefont {Giovannetti},\ and\ \citenamefont
  {Calarco}}]{Murphy2010}%
  \BibitemOpen
  \bibfield  {author} {\bibinfo {author} {\bibfnamefont {M.}~\bibnamefont
  {Murphy}}, \bibinfo {author} {\bibfnamefont {S.}~\bibnamefont {Montangero}},
  \bibinfo {author} {\bibfnamefont {V.}~\bibnamefont {Giovannetti}}, \ and\
  \bibinfo {author} {\bibfnamefont {T.}~\bibnamefont {Calarco}},\ }\href
  {\doibase 10.1103/PhysRevA.82.022318} {\bibfield  {journal} {\bibinfo
  {journal} {Phys. Rev. A}\ }\textbf {\bibinfo {volume} {82}},\ \bibinfo
  {pages} {022318} (\bibinfo {year} {2010})}\BibitemShut {NoStop}%
\bibitem [{\citenamefont {Wildberger}\ \emph {et~al.}(1995)\citenamefont
  {Wildberger}, \citenamefont {Stepanyuk}, \citenamefont {Lang}, \citenamefont
  {Zeller},\ and\ \citenamefont {Dederichs}}]{Wildberger1995}%
  \BibitemOpen
  \bibfield  {author} {\bibinfo {author} {\bibfnamefont {K.}~\bibnamefont
  {Wildberger}}, \bibinfo {author} {\bibfnamefont {V.~S.}\ \bibnamefont
  {Stepanyuk}}, \bibinfo {author} {\bibfnamefont {P.}~\bibnamefont {Lang}},
  \bibinfo {author} {\bibfnamefont {R.}~\bibnamefont {Zeller}}, \ and\ \bibinfo
  {author} {\bibfnamefont {P.~H.}\ \bibnamefont {Dederichs}},\ }\href {\doibase
  10.1103/PhysRevLett.75.509} {\bibfield  {journal} {\bibinfo  {journal} {Phys.
  Rev. Lett.}\ }\textbf {\bibinfo {volume} {75}},\ \bibinfo {pages} {509}
  (\bibinfo {year} {1995})}\BibitemShut {NoStop}%
\bibitem [{\citenamefont {Zeller}\ \emph {et~al.}(1995)\citenamefont {Zeller},
  \citenamefont {Dederichs}, \citenamefont {{\'U}jfalussy}, \citenamefont
  {Szunyogh},\ and\ \citenamefont {Weinberger}}]{Zeller1995}%
  \BibitemOpen
  \bibfield  {author} {\bibinfo {author} {\bibfnamefont {R.}~\bibnamefont
  {Zeller}}, \bibinfo {author} {\bibfnamefont {P.~H.}\ \bibnamefont
  {Dederichs}}, \bibinfo {author} {\bibfnamefont {B.}~\bibnamefont
  {{\'U}jfalussy}}, \bibinfo {author} {\bibfnamefont {L.}~\bibnamefont
  {Szunyogh}}, \ and\ \bibinfo {author} {\bibfnamefont {P.}~\bibnamefont
  {Weinberger}},\ }\href {\doibase 10.1103/PhysRevB.52.8807} {\bibfield
  {journal} {\bibinfo  {journal} {Phys. Rev. B}\ }\textbf {\bibinfo {volume}
  {52}},\ \bibinfo {pages} {8807} (\bibinfo {year} {1995})}\BibitemShut
  {NoStop}%
\bibitem [{\citenamefont {Stepanyuk}\ \emph {et~al.}(2006)\citenamefont
  {Stepanyuk}, \citenamefont {Negulyaev}, \citenamefont {Niebergall},
  \citenamefont {Longo},\ and\ \citenamefont {Bruno}}]{Stepanyuk2006}%
  \BibitemOpen
  \bibfield  {author} {\bibinfo {author} {\bibfnamefont {V.~S.}\ \bibnamefont
  {Stepanyuk}}, \bibinfo {author} {\bibfnamefont {N.~N.}\ \bibnamefont
  {Negulyaev}}, \bibinfo {author} {\bibfnamefont {L.}~\bibnamefont
  {Niebergall}}, \bibinfo {author} {\bibfnamefont {R.~C.}\ \bibnamefont
  {Longo}}, \ and\ \bibinfo {author} {\bibfnamefont {P.}~\bibnamefont
  {Bruno}},\ }\href {\doibase 10.1103/PhysRevLett.97.186403} {\bibfield
  {journal} {\bibinfo  {journal} {Phys. Rev. Lett.}\ }\textbf {\bibinfo
  {volume} {97}},\ \bibinfo {pages} {186403} (\bibinfo {year}
  {2006})}\BibitemShut {NoStop}%
\bibitem [{\citenamefont {Negulyaev}\ \emph {et~al.}(2008)\citenamefont
  {Negulyaev}, \citenamefont {Stepanyuk}, \citenamefont {Niebergall},
  \citenamefont {Bruno}, \citenamefont {Hergert}, \citenamefont {Repp},
  \citenamefont {Rieder},\ and\ \citenamefont {Meyer}}]{Negulyaev2008}%
  \BibitemOpen
  \bibfield  {author} {\bibinfo {author} {\bibfnamefont {N.~N.}\ \bibnamefont
  {Negulyaev}}, \bibinfo {author} {\bibfnamefont {V.~S.}\ \bibnamefont
  {Stepanyuk}}, \bibinfo {author} {\bibfnamefont {L.}~\bibnamefont
  {Niebergall}}, \bibinfo {author} {\bibfnamefont {P.}~\bibnamefont {Bruno}},
  \bibinfo {author} {\bibfnamefont {W.}~\bibnamefont {Hergert}}, \bibinfo
  {author} {\bibfnamefont {J.}~\bibnamefont {Repp}}, \bibinfo {author}
  {\bibfnamefont {K.-H.}\ \bibnamefont {Rieder}}, \ and\ \bibinfo {author}
  {\bibfnamefont {G.}~\bibnamefont {Meyer}},\ }\href {\doibase
  10.1103/PhysRevLett.101.226601} {\bibfield  {journal} {\bibinfo  {journal}
  {Phys. Rev. Lett.}\ }\textbf {\bibinfo {volume} {101}},\ \bibinfo {eid}
  {226601} (\bibinfo {year} {2008})}\BibitemShut {NoStop}%
\bibitem [{\citenamefont {Lagoute}\ \emph {et~al.}(2007)\citenamefont
  {Lagoute}, \citenamefont {Nacci},\ and\ \citenamefont
  {F{\"o}lsch}}]{Lagoute2007}%
  \BibitemOpen
  \bibfield  {author} {\bibinfo {author} {\bibfnamefont {J.}~\bibnamefont
  {Lagoute}}, \bibinfo {author} {\bibfnamefont {C.}~\bibnamefont {Nacci}}, \
  and\ \bibinfo {author} {\bibfnamefont {S.}~\bibnamefont {F{\"o}lsch}},\
  }\href {\doibase 10.1103/PhysRevLett.98.146804} {\bibfield  {journal}
  {\bibinfo  {journal} {Phys. Rev. Lett.}\ }\textbf {\bibinfo {volume} {98}},\
  \bibinfo {eid} {146804} (\bibinfo {year} {2007})}\BibitemShut {NoStop}%
\bibitem [{\citenamefont {Shiraki}\ \emph {et~al.}(2004)\citenamefont
  {Shiraki}, \citenamefont {Fujisawa}, \citenamefont {Nantoh},\ and\
  \citenamefont {Kawai}}]{Shiraki2004}%
  \BibitemOpen
  \bibfield  {author} {\bibinfo {author} {\bibfnamefont {S.}~\bibnamefont
  {Shiraki}}, \bibinfo {author} {\bibfnamefont {H.}~\bibnamefont {Fujisawa}},
  \bibinfo {author} {\bibfnamefont {M.}~\bibnamefont {Nantoh}}, \ and\ \bibinfo
  {author} {\bibfnamefont {M.}~\bibnamefont {Kawai}},\ }\href {\doibase
  10.1103/PhysRevLett.92.096102} {\bibfield  {journal} {\bibinfo  {journal}
  {Phys. Rev. Lett.}\ }\textbf {\bibinfo {volume} {92}},\ \bibinfo {pages}
  {096102} (\bibinfo {year} {2004})}\BibitemShut {NoStop}%
\bibitem [{\citenamefont {Hirjibehedin}\ \emph {et~al.}(2006)\citenamefont
  {Hirjibehedin}, \citenamefont {Lutz},\ and\ \citenamefont
  {Heinrich}}]{Hirjibehedin2006}%
  \BibitemOpen
  \bibfield  {author} {\bibinfo {author} {\bibfnamefont {C.~F.}\ \bibnamefont
  {Hirjibehedin}}, \bibinfo {author} {\bibfnamefont {C.~P.}\ \bibnamefont
  {Lutz}}, \ and\ \bibinfo {author} {\bibfnamefont {A.~J.}\ \bibnamefont
  {Heinrich}},\ }\href {\doibase 10.1126/science.1125398} {\bibfield  {journal}
  {\bibinfo  {journal} {Science}\ }\textbf {\bibinfo {volume} {312}},\ \bibinfo
  {pages} {1021} (\bibinfo {year} {2006})}\BibitemShut {NoStop}%
\bibitem [{\citenamefont {D\'\i{}az-Tendero}\ \emph
  {et~al.}(2009{\natexlab{a}})\citenamefont {D\'\i{}az-Tendero}, \citenamefont
  {Borisov},\ and\ \citenamefont {Gauyacq}}]{Diaz-Tendero2009prl}%
  \BibitemOpen
  \bibfield  {author} {\bibinfo {author} {\bibfnamefont {S.}~\bibnamefont
  {D\'\i{}az-Tendero}}, \bibinfo {author} {\bibfnamefont {A.~G.}\ \bibnamefont
  {Borisov}}, \ and\ \bibinfo {author} {\bibfnamefont {J.-P.}\ \bibnamefont
  {Gauyacq}},\ }\href {\doibase 10.1103/PhysRevLett.102.166807} {\bibfield
  {journal} {\bibinfo  {journal} {Phys. Rev. Lett.}\ }\textbf {\bibinfo
  {volume} {102}},\ \bibinfo {pages} {166807} (\bibinfo {year}
  {2009}{\natexlab{a}})}\BibitemShut {NoStop}%
\bibitem [{\citenamefont {D\'\i{}az-Tendero}\ \emph
  {et~al.}(2009{\natexlab{b}})\citenamefont {D\'\i{}az-Tendero}, \citenamefont
  {Olsson}, \citenamefont {Borisov},\ and\ \citenamefont
  {Gauyacq}}]{Diaz-Tendero2009}%
  \BibitemOpen
  \bibfield  {author} {\bibinfo {author} {\bibfnamefont {S.}~\bibnamefont
  {D\'\i{}az-Tendero}}, \bibinfo {author} {\bibfnamefont {F.~E.}\ \bibnamefont
  {Olsson}}, \bibinfo {author} {\bibfnamefont {A.~G.}\ \bibnamefont {Borisov}},
  \ and\ \bibinfo {author} {\bibfnamefont {J.-P.}\ \bibnamefont {Gauyacq}},\
  }\href {\doibase 10.1103/PhysRevB.79.115438} {\bibfield  {journal} {\bibinfo
  {journal} {Phys. Rev. B}\ }\textbf {\bibinfo {volume} {79}},\ \bibinfo
  {pages} {115438} (\bibinfo {year} {2009}{\natexlab{b}})}\BibitemShut
  {NoStop}%
\bibitem [{\citenamefont {Brovko}\ \emph {et~al.}(2008)\citenamefont {Brovko},
  \citenamefont {Ignatiev}, \citenamefont {Stepanyuk},\ and\ \citenamefont
  {Bruno}}]{Brovko2008prl}%
  \BibitemOpen
  \bibfield  {author} {\bibinfo {author} {\bibfnamefont {O.~O.}\ \bibnamefont
  {Brovko}}, \bibinfo {author} {\bibfnamefont {P.~A.}\ \bibnamefont
  {Ignatiev}}, \bibinfo {author} {\bibfnamefont {V.~S.}\ \bibnamefont
  {Stepanyuk}}, \ and\ \bibinfo {author} {\bibfnamefont {P.}~\bibnamefont
  {Bruno}},\ }\href {\doibase 10.1103/PhysRevLett.101.036809} {\bibfield
  {journal} {\bibinfo  {journal} {Phys. Rev. Lett.}\ }\textbf {\bibinfo
  {volume} {101}},\ \bibinfo {eid} {036809} (\bibinfo {year}
  {2008})}\BibitemShut {NoStop}%
\end{thebibliography}

\end{document}